\documentclass[12pt, referee]{aastex}
\usepackage[dvips]{color}
\newcommand{\re}{\textcolor[rgb]{0,0,0}}

\newcommand{\bl}{\textcolor[rgb]{0,0,0}}

\newcommand{\ma}{\textcolor[rgb]{0.0,0,0.0}}
\newcommand{\gr}{\textcolor[rgb]{0,0.0,0}}

\def\gtsima{$\;\buildrel > \over \sim \;$}
\def\simgt{\lower.5ex \hbox{\gtsima}}
\def\ltsima{$\;\buildrel < \over \sim \;$}
\def\simlt{\lower.5ex \hbox{\ltsima}}

\begin{document}

\title{Dark Matter that Interacts with Baryons: Experimental Limits on the Interaction Cross-section for 27 Atomic Nuclei, and Resultant Constraints on the Particle Properties}

\author{David A. Neufeld\altaffilmark{1} and Daniel J. Brach-Neufeld}

\altaffiltext{1}{Department of Physics \& Astronomy, Johns Hopkins University, Baltimore, MD 20218, USA}

\begin{abstract}

To constrain the properties of dark matter (DM) that interacts with nucleons, we have conducted an experimental search for any anomalous heating of ordinary baryonic matter at 77 K. Our tabletop experiment is motivated by the possibility (discussed in a previous paper) that DM particles with masses in the $\sim 1 - 2 m_{\rm p}$ range could be captured by and concentrated within the Earth. For suitable parameters, this phenomenon could lead to a substantial density ($\sim 10^{14}\,\rm cm^{-3}$) of thermalized (300 K) DM particles at Earth's surface that would heat cooler baryonic matter. Our experiment involves precise differential measurements of the evaporation rate of liquid nitrogen in a storage dewar within which various materials are immersed. The results revealed no statistically-significant detections of heating in the 27 elements with molar fractions $\simgt 10^{-5}$ in Earth's crust.  For material with the average composition of Earth's crust, our measurements imply a $3 \sigma$ upper limit of $1.32 \times 10^{-27}\, n_{\rm 14}^{-1} (m_{\rm DM}/2\,m_{\rm p})^{-1/2}\,\rm cm^2$ on the mean cross-section for scattering with thermal HIDM at 300 K, where $10^{14}\,n_{14} \, \rm cm^{-3}$ is the particle density at Earth's surface. In combination with a lower limit on the scattering cross-section, obtained from a consideration of the heat flow through the Earth's crust (Neufeld, Farrar \& McKee 2018), our experiment places an upper limit of $1.6 \times 10^{13}\,\rm cm^{-3}$ on the density of DM at the Earth's surface.  This in turn, significantly constrains the properties for any DM candidate that interacts with baryons.

\end{abstract}


\section{Introduction}

In recent years, several studies have considered the possibility that dark matter (DM) might interact significantly with ordinary baryonic matter through non-gravitational forces, and the possible effects of such interactions have been evaluated in multiple astrophysical contexts.  These considerations have been
discussed in the introductory section of a recent paper (Neufeld, Farrar \& McKee 2018; hereafter NFM18) and will not be repeated here.

One recently-proposed candidate for a hadronically-interacting DM (hereafter HIDM) particle is the neutral, spinless sexaquark (Farrar 2017), $uuddss$, which would be absolutely stable if its mass were less than $2\,(m_{\rm p}+ m_{\rm e})$.  As it moves through the Galactic disk, the Earth would intercept such HIDM, 
which could be captured, thermalized, retained, and thereby greatly concentrated relative to their density in the disk (NFM18).  
The particle number density at the Earth's surface, $n_{\rm DM}(R_\earth)$, depends strongly on the particle mass, $m_{\rm DM}$, and may exceed $10^{14}\, \rm cm^{-3}$ for masses in the $1 - 2$~m$_p$ range.  For larger masses, the density distribution within the Earth becomes increasingly centrally-condensed, and $n_{\rm DM}(R_\earth)$ therefore decreases with increasing $m_{\rm DM}$; while for smaller $m_{\rm DM}$, 
$n_{\rm DM}(R_\earth)$ drops rapidly with decreasing $m_{\rm DM}$, because the evaporation (i.e.\ ``Jeans loss") of captured HIDM  becomes significant over the lifetime of the Earth.  The latter effect depends sensitively on the temperature, $T_{\rm LSS}$, at the ``last scattering surface" (LSS) where an escaping HIDM particle last scatters with baryonic matter in the Earth's crust or atmosphere, which depends in turn on the scattering cross-section.  Results for $n_{\rm DM}(R_\earth)$ were presented by NFM18 (their Figure 6) for $m_{DM}$ in the $0.5 - 10\,m_{\rm p}$ range and scattering cross-sections in the $10^{-30}$ to $10^{-20} \, \rm cm^2$ range.  \bl{For $m_{\rm DM} = 1 - 2\,m_{\rm p}$}, the natural \bl{range} for a stable sexaquark, the surface HIDM particle density \bl{can be as large as $10^{14}\, \rm cm^{-3}$} provided that the average scattering cross-section for atoms in the Earth's crust is larger than $5 \times 10^{-29}\, \rm cm^2$.  For these parameters, the LSS lies at a depth less than $3\, \rm km$ below the Earth's surface, where the typical temperature, $T_{\rm LSS}$, is below 400~K.  However, for smaller values of the cross-section, the LSS sinks further into the crust where the temperature is higher, particle evaporation becomes significant, and $n_{\rm DM}(R_\earth)$ falls rapidly. 

For particle masses, $m_{\rm DM}$ in the 0.6 -- 6~$m_{\rm p}$ range, NFM18 derived quantitative constraints on the interaction cross-sections for scattering with ordinary baryonic matter.  These constraints were placed by four considerations: 

\noindent (1) The $\sim 100$~hr lifetime of the relativistic proton beam at the Large Hadron collider (LHC) places an  upper limit on the cross-section for inelastic collisions between stationary HIDM and 6.5 TeV protons.

\noindent (2) The orbital decay of spacecraft in low earth orbit (LEO) --  and the observed orbital decay rate for the Hubble Space Telescope (HST) in particular -- place an upper limit on orbital drag caused by HIDM, thereby providing upper limits on the scattering cross-sections for the major constituents of HST (e.g.\ Al, Si, O).

\noindent (3) The observed vaporization rates of cryogenic liquids in well-insulated storage dewars (e.g. liquid helium, hydrogen, oxygen, nitrogen, and argon) place limits on the rate at which thermal HIDM in the laboratory heat the contents of a dewar, providing upper limits on the scattering cross-sections for He, H, O, N, Ar.  These constraints arise because collisions between HIDM at the temperature of the laboratory and nuclei at cryogenic temperatures transfer an amount of heat that is proportional to the temperature difference (see equation (1) in section 4 below.)

\noindent (4) The thermal conductivity of the Earth's crust, which is constrained by measurements of temperature gradients in boreholes, places an upper limit on any anomalous conductivity associated with HIDM.  This, in turn, provides an upper limit on the mean-free path for HIDM within the crust, and thereby a lower limit on the average scattering cross-section for abundant nuclei in the crust.  \bl{We emphasize that this consideration leads to a lower limit on the cross-section whereas the other considerations
yield upper limits.}  For a particle density of $10^{14}\,\rm{cm}^{3}$ and a particle mass of $2\, m_{\rm p}$, this lower limit on the scattering cross-section was found to greatly {\it exceed} the upper limits obtained for He and H from the consideration of liquid cryogens, and to marginally exceed those obtained for N, O and Ar.   

Given that the cross-sections for the scattering of HIDM by atomic nuclei may depend strongly and non-monotonically on the nucleon number (X.~Xu \& G~.R.~Farrar 2019, in preparation), as is indeed the case for scattering of neutrons by atomic nuclei, the limits obtained by NFM18 apply to specific nuclei.  In the case of the third consideration discussed above, the results presented by NFM18 were limited to five nuclei present in cryogenic liquids for which dewar boil-off rates were available from published literature or from  specification sheets provided by dewar manufacturers.  Because the crust contains many abundant nuclei  (e.g. Ca, Na, K) for which the first three considerations did not provide upper limits on the scattering cross-section with HIDM, the lower limits obtained from consideration (4) are not necessarily in conflict with those upper limits. 

In the \gr{simple and inexpensive\footnote{\gr{The total cost of the necessary equipment and supplies was $\sim \$7\rm K$ (USD)}} tabletop}
experiment reported on here, we have improved the upper limits reported by NFM18 for the scattering of HIDM by O and N nuclei, and greatly \bl{expanded} the number of nuclei for which such limits are now available, \bl{with the goal of placing robust
limits on the mean cross-section for material of crustal composition.}  
As described in Section 2 below, our experiment \bl{searched} for any anomalous heating of samples of different solid materials containing a variety of nuclei that are immersed in liquid nitrogen (LN2) within storage dewars.  \gr{Our method is 
complementary to other direct detection experiments -- e.g.\ CDMS (Abusaidi et al.\ 2000; Abrams et al.\ 2002), DAMIC (Aguilar-Arevalo et al.\ 2016), COSINUS (Angloher et al.\ 2017), LUX (Akerib et al.\ 2017), XENON (Aprile et al.\ 2017), PandaX (Cui et al.\ 2017) -- which can set much stronger limits on the cross-sections for 
200~km~s$^{-1}$ DM particles but are insensitive to DM at room temperature thermal energies (and therefore blind to
HIDM particles that are stopped in the overburden above the detector).  Moreover, 
other direct detection experiments are typically sensitive only to interactions with a specific element
(usually Si, Ge, O or Al; see Hooper \& McDermott 2018, their Table 1)}.
The results of our experiment are presented in Section 3, and their implications are discussed in Section 4.

\section{Experimental set-up and procedure}

Our experiment involves the use of a precision balance to measure the vaporization of LN2 from \bl{three} 
continuously-vented 6-liter storage dewars in a temperature-controlled environment.  
The balance, a SAW-T scale manufactured by the Arlyn Scales company, permits masses up to 11.4~kg to be measured with a precision of 0.1~g, and was calibrated repeatedly with the use of a 10~kg Class M1 calibration mass.  These calibrations indicated that the balance can be used to obtain mass determinations that have an accuracy of 0.23~g (1$\,\sigma$).
The dewars, \bl{model USS-LNT00002}, were manufactured by US Solid, and were found to exhibit LN2 evaporation rates in the range $\sim 100 - 110 \, \rm g\, day^{-1}$ for an ambient temperature of 24~$^0$C.  Given a 
latent heat of vaporization $L_{\rm vap} = 199 \, \rm J\,g^{-1}$ at 1 atm pressure, this evaporation rate corresponds to a dewar hold time of $\sim 45$~days and a heat leak of $\sim 250$~mW.  

During a given experimental run of typical duration 6 days, the three dewars were weighed roughly once per day, allowing the average vaporization rate to be determined.   For each dewar, week-long runs alternated between control runs with only LN2 inside the dewar, and sample runs in which $1 - 2$~kg of \bl{sample} material was completely immersed in the LN2.   Thus, our experiment involves a differential measurement in which the heating rates are compared with and without immersed samples.  The control runs were staggered with the sample runs, so that in any given time \bl{at least} one dewar was undergoing a control run.  \bl{This procedure provides some sensitivity to the influence of unidentified environmental effects on the boil-off rates, since the control run may be compared with others conducted for that same dewar (e.g. that conducted two weeks previously)}.

For sample materials that could be poured out of the dewar easily by simply inverting it -- such as low-carbon steel bearings, silicon lumps, and magnesium shot -- the sample was immersed loose within the dewar.  For fine granular materials, however, we packed the sample within up to ten aluminum vials, each of capacity $\sim 120$~ml.  A pair of \bl{linked} steel key rings was attached to each vial, permitting it to be retrieved at the end of each run using a magnet on a telescoping rod.  When initially immersed in the LN2, the pressure in the vials drops rapidly, which can result in the entry of a small amount of LN2 around the screw-on vial caps.  Upon removal at the end of a run, this LN2 heats up rapidly, presenting a risk of explosion, which we mitigated by creating safety valves within the cap of each vial.  These were constructed by drilling a small hole in each cap and covering the cap interior with aluminum foil that will break if the pressure build-up is too large.  

The ambient temperature was controlled to within roughly 0.3$^0$C, with the use of a heater and a thermostat, and was monitored with bluetooth logging thermometers (Onset models MX-\rm{1101} and MX-\rm{2303}).  
The latter provided a precision of 0.01$^0$C, an absolute accuracy of 0.1$^0$C, and a sampling interval of 1 minute, and were used to monitor the air temperature and the temperature at two locations on the outer surface of each dewar by means of thermistors attached with aluminum tape.  By performing runs both at elevated ambient temperatures $\sim 30^0$C \bl{and at reduced temperatures $\sim 20^0$C,} we determined the dependence of the evaporation rate on the dewar surface temperature for each dewar.  Temperature coefficients in the range \bl{$0.81$ to $0.90 \, \rm g\, day^{-1}\, K^{-1}$} were thereby measured \bl{separately} 
for the three dewars, and were then used to correct for small variations in the dewar surface temperatures (that resulted from the imperfect control of temperature.)
Relative humidity and pressure were also logged (\bl{but not controlled}), using the Onset MX-\rm{1101} and a pressure monitor (GCDC Model B1100-2).  

\begin{figure}
\includegraphics[width=16 cm]{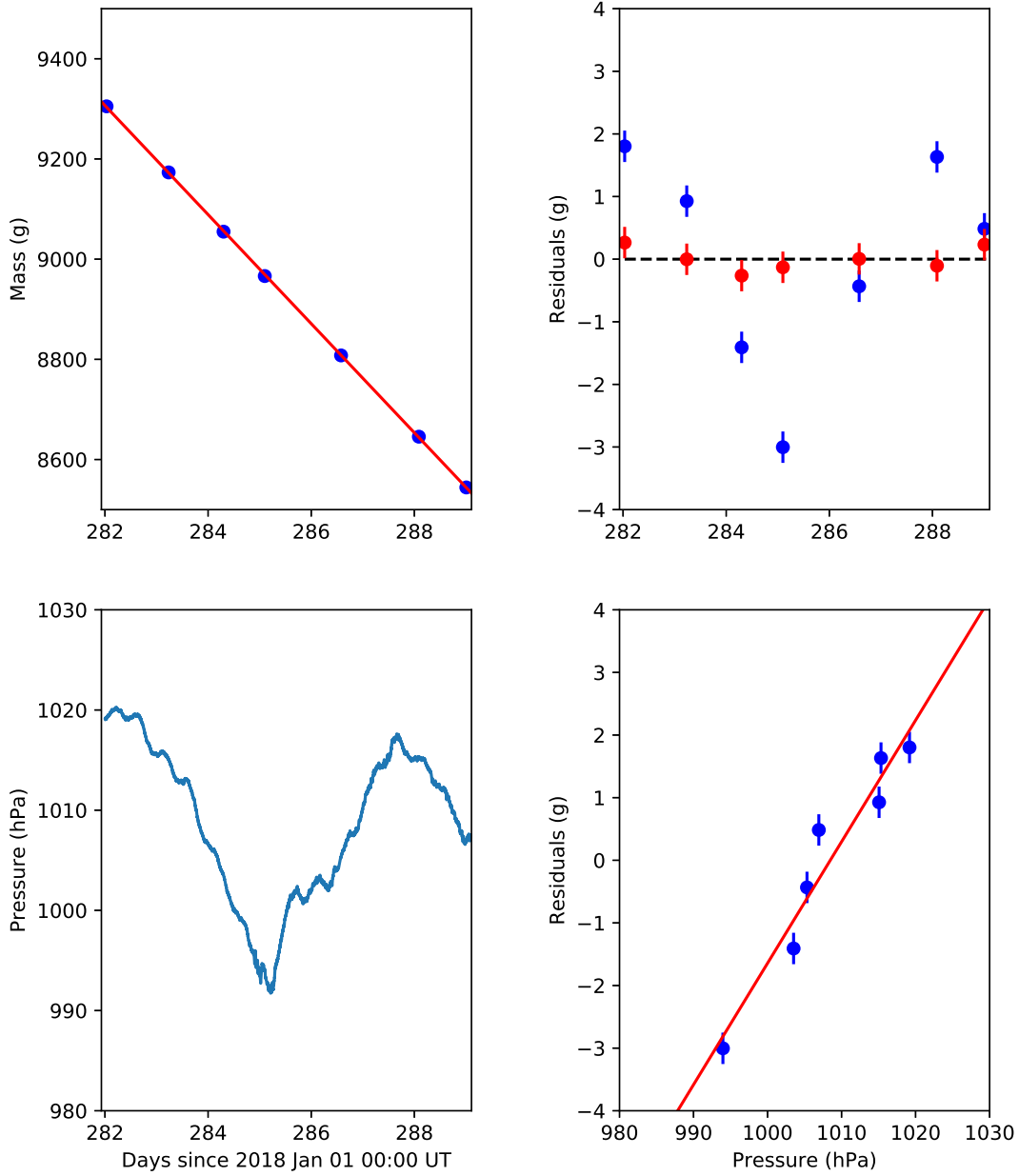}
\caption{Example results obtained for a control run of Dewar 3.  The error bars in the upper left panel are much smaller than the size of the circles plotted.}
\end{figure}

\section{Results}

Figure 1 presents example results obtained for a control run.  The top left panel shows the mass of the dewar and its contents as a function of time (blue points).  The best-fit linear regression is shown by the red curve.  The blue points in the top right panel show the mass residuals from that fit, which are found to be linearly correlated (bottom right panel) with the ambient pressure (shown in hPa in the bottom left panel.)  The mass residuals show a linear dependence\footnote{\bl{This correlation results from the dependence of the LN2 boiling point on pressure.  If the pressure increases by an amount $dp$, the boiling point of the LN2 is elevated by $dT_b$; 
additional heat $dQ = C dT_b,$ must therefore leak 
into the dewar before the evaporation can resume, where $C$ is the heat capacity of the system (including both 
the contents of the dewar and the inner aluminum vessel that is at the temperature of the cryogen).
This effect leads to a positive mass residual, $dM = dQ / L_{\rm vap} = C dT_b / L_{\rm vap}.$ 
The change in boiling point is related to the change in pressure by the Clausius-Clapeyron relation: 
$dT_b/dp = T_b \Delta v / L_{\rm vap},$ where $\Delta v$ is the change in specific volume associated with
the phase change.  The specific volume in the liquid phase is negligible relative to that in the gas-phase, 
so $\Delta v$ may be approximated by $kT_b/m_{\rm N2}$, where $m_{\rm N2} = 28$~a.m.u. is the mass of an N$_2$ molecule.
Combining these expressions, we obtain
$${dM \over  dp} =  {kT_b^2 C \over m_{\rm N2} L_{\rm vap}^2  p }. $$
At 1 atm pressure, the yields $dM / dp = 0.0442\,C_3\rm \, g \,hPa^{-1},$ where $C_3=C/(\rm 10^3\,J\,K^{-1}).$
For the example case shown in Figure 1, the observed value of $dM/dp$ implies a system heat capacity of 
$3800\,\rm J\,K^{-1}$.}} on the pressure with a 
slope $dM/dp = 0.17\rm \, g \, hPa^{-1},$  where hPa denotes the hectoPascal (also known as the millibar: 1~hPa = $10^2 \rm \, Pa$ = 1 millibar). 
ß

After correction for the effects of varying pressure and temperature, the mass residuals reduce to the red points shown in the top left panel.  
\bl{In this run, the r.m.s. of the residuals was 180~mg, after correction for the varying pressure, and the standard error 
on the evaporation rate was 38~$\rm \, mg\, day^{-1}$, corresponding to a total heating rate of $89 \rm \, \mu W$ for 
the contents of the dewar.}  The $\chi^2$ per degree of freedom (DOF) for this case was 1.06, given 4 DOF resulting from seven data points with 
three fitted parameters (starting mass, vaporization rate, and $dM/dp$).
This goodness-of-fit implies that after correction for the pressure and temperature dependences the remaining residuals are entirely consistent with the accuracy of the mass measurements; thus, there is no evidence that any additional environmental factor is affecting the results.

\begin{figure}
\includegraphics[width=15 cm]{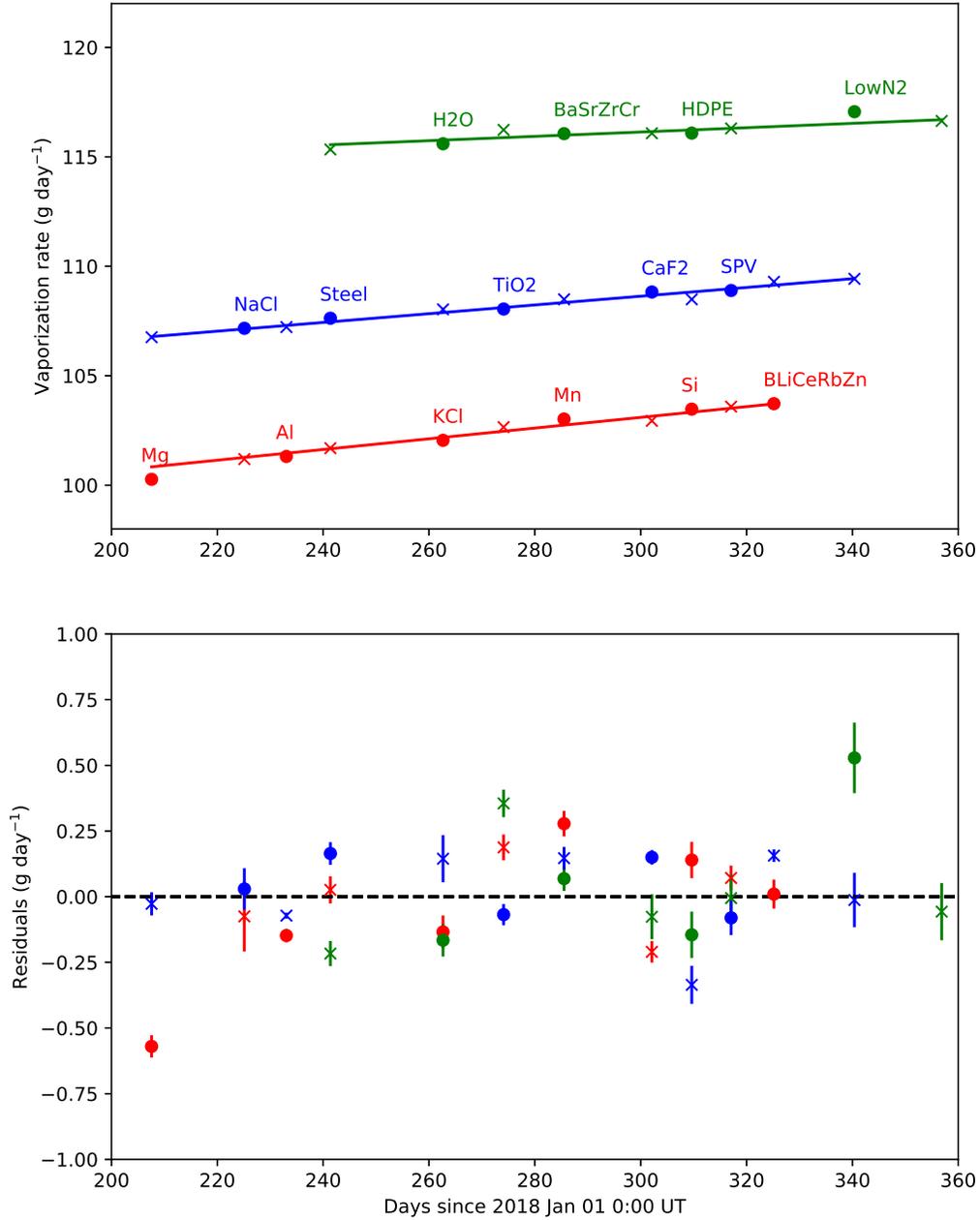}
\caption{Vaporization rates in different runs for Dewar 1 (red), Dewar 2 (green), and Dewar 3 (blue).   The error bars in the upper panel are typically smaller than the size of the circles plotted.}
\end{figure}

The upper panel of Figure 2 shows the results obtained for the various experimental runs, as a function of time (i.e.\ days since 2018 Jan 01 00:00 UT) at the midpoint of each run.  Different colors denote the three different dewars.  \bl{Crosses refer to control runs, and filled circles refer to sample runs, with
the labels indicating the materials being tested.}  The obvious upwards trend in the vaporization rates for all three dewars apparently represents a slow deterioration in the thermal insulation, which we speculate is caused by the leakage of air into the evacuated space between the outer and inner vessels that comprise the dewar.  
The solid lines show the best-fit linear regression to the results we obtained in the control runs.  
The lower panel of Figure 2 shows the residuals from these linear fits to the long-term variation of the
boil-off rate.  Here, the linear fits yield reduced-$\chi^2$ values of 15, 23, and 24 respectively for dewars 1, 2, and 3, 
implying that there are residual, statistically-significant run-to-run variations that are not accounted for 
by a simple linear deterioration in the dewar heat leak.  The typical magnitude of these variations 
is a factor 4 -- 5 times the typical measurement uncertainty for a single run.

We have adopted the standard deviation of these variations 
from the linear fit as our estimate of the  1~$\sigma$ uncertainties on any anomalous boil-off rate.  
These standard deviations are 0.173, 0.246 and 0.193 g~day$^{-1}$ 
respectively for dewars 1, 2, and 3, corresponding to total heating rates of 0.40, 0.57 and 0.45 mW.

\begin{table}
\scriptsize
\caption{\ma{Anomalous Heating rates}}
\vskip 0.1 true in
\begin{tabular}{lcccrrcr}

\hline
Material			& Start time 	& End time 	& Run ID/ 	& Mass 		& Number of & Nucleus 	& Molar heating \\
 				&  \multicolumn{2}{l}{(Days since 2018 Jan 1 0:00 UT)}	& Dewar ID	& (g)$\phantom{0}$		& moles$\phantom{00}$	  &            & heating rate \\
 				&			&			& 			&			& 		  &            & ($\rm \mu W \, mol^{-1}$) \\
\hline
Mg              &   204.6  &  210.6  & A/1 & 2192.0 & 90.1 &  Mg & $   -14.6 \pm     4.4  \phantom{00}$\\
NaCl            &   221.6  &  228.6  & B/3 & 1653.0 & 28.0 &  Na & $    +2.4 \pm    15.8  \phantom{0}$\\
...             &   221.6  &  228.6  & B/3 & 1653.0 & 28.0 &  Cl & $    +2.4 \pm    15.8  \phantom{0}$\\
Al              &   229.5  &  236.6  & C/1 & 1860.0 & 68.9 &  Al & $    -4.9 \pm     5.8  \phantom{00}$\\
Steel           &   237.9  &  244.8  & D/3 & 2041.0 & 36.2 &  Fe & $   +10.5 \pm    12.3  \phantom{0}$\\
KCl             &   256.7  &  268.6  & E/1 & 1192.0 & 15.8 &  K & $   -19.5 \pm    25.3  \phantom{0}$\\
H$_2$O          &   256.7  &  268.6  & E/2 & 997.3 & 55.3 &  O & $    -6.9 \pm    10.3  \phantom{0}$\\
...             &   256.7  &  268.6  & E/2 & 997.3 & 110.6 &  H & $    -3.5 \pm     5.1  \phantom{00}$\\
TiO$_2$         &   271.1  &  277.1  & F/3 & 1244.0 & 15.4 &  Ti & $   -10.2 \pm    28.8  \phantom{0}$\\
MnCO$_3$        &   282.0  &  289.0  & G/1 & 1686.0 & 14.5 &  Mn & $   +44.1 \pm    27.6  \phantom{0}$\\
ZrSiO$_4$       &   282.0  &  289.0  & G/2 & 379.1 & 2.0 &  Zr & $   +77.2 \pm   277.2  $\\
Cr$_2$O$_3$     &   282.0  &  289.0  & G/2 & 371.7 & 4.8 &  Cr & $   +32.7 \pm   117.2  $\\
SrCO$_3$        &   282.0  &  289.0  & G/2 & 522.4 & 3.5 &  Sr & $   +45.1 \pm   162.0  $\\
BaCO$_3$        &   282.0  &  289.0  & G/2 & 559.6 & 2.8 &  Ba & $   +56.3 \pm   202.1  $\\
CaF$_2$         &   299.1  &  305.1  & H/3 & 1859.6 & 23.6 &  Ca & $   +14.6 \pm    18.8  \phantom{0}$\\
...             &   299.1  &  305.1  & H/3 & 1859.6 & 47.2 &  F & $    +7.3 \pm     9.4  \phantom{00}$\\
Si              &   306.6  &  312.7  & I/1 & 2000.8 & 70.2 &  Si & $    +4.6 \pm     5.7  \phantom{00}$\\
HDPE            &   306.6  &  312.7  & I/2 & 1480.4 & 104.5 &  C & $    -3.2 \pm     5.4  \phantom{00}$\\
V$_2$O$_5$      &   314.1  &  320.0  & J/3 & 258.1 & 2.8 &  V & $   -66.0 \pm   157.9  $\\
NaH$_2$PO$_4$   &   314.1  &  320.0  & J/3 & 682.8 & 5.7 &  P & $   -32.6 \pm    78.0  \phantom{0}$\\
S               &   314.1  &  320.0  & J/3 & 600.4 & 18.7 &  S & $    -9.9 \pm    23.7  \phantom{0}$\\
H$_3$BO$_3$     &   321.7  &  328.7  & K/1 & 251.4 & 4.0 &  B & $    +5.6 \pm    99.5  \phantom{0}$\\
RbCl            &   321.7  &  328.7  & K/1 & 58.7 & 0.5 &  Rb & $   +47.1 \pm   833.1  $\\
Zn              &   321.7  &  328.7  & K/1 & 782.7 & 11.9 &  Zn & $    +1.9 \pm    33.8  \phantom{0}$\\
Li$_2$CO$_3$    &   321.7  &  328.7  & K/1 & 181.6 & 4.9 &  Li & $    +4.6 \pm    82.3  \phantom{0}$\\
CeO$_2$         &   321.7  &  328.7  & K/1 & 542.4 & 3.1 &  Ce & $    +7.2 \pm   128.3  $\\
N$_2$           &   336.8  &  344.0  & L/2 & -2600.0 & 183.8 &  N & $    -6.6 \pm     3.1  \phantom{00}$\\

\hline 
\end{tabular}
\end{table}

In Table 1, we list the materials tested, the start and end times for each run, the dewar used for each run, 
the mass of material, \bl{the corresponding number of moles},
the elements thereby tested, and the molar heating rates that were measured.  For most elements with molar 
fractions $\le 10^{-3}$ in the crust, multiple materials were tested within a single dewar during a single run. 
In cases where an element was tested in multiple runs (e.g. for Cl, which is present in both NaCl and KCl), 
only the stronger limit is listed.  \bl{The purity of all tested materials exceeded 98.5$\%$.}  \gr{Our determination of
the heating rate for N was obtained from a run in which the LN2 volume was greatly reduced relative to that in all
the control runs; the effective sample mass is therefore negative.}   The level of the LN2 was considerably lower in 
this run than in any other run.  Because
the heat leak in the dewar is only expected to be a decreasing function the LN2 depth -- if indeed there
is any measurable dependence at all --  this run yields a lower limit on the heating rate for N.   
\bl{Table 1 shows that none of the sample runs yielded a statistically-significant departure from a 
null result.}\footnote{\bl{Run A (Mg shot), yielded a negative deviation of 3.39$\sigma$, the most significant
deviation in the set of 15 sample runs.  The probability of a deviation of that absolute magnitude 
(or larger) occurring by chance in any single run is $\sim 4\%$ (Student t-distribution with $N - 2$ degrees of freedom, 
where $N= 5$ is the number of control runs from which the linear fit in Figure 2 was obtained.)  \gr{The probability of
one or more such deviations occurring among the 15 sample runs is therefore 46\%.}}} 
\bl{For the ten most abundant elements in the crust, the 1$\sigma$ uncertainties in the molar heating rates 
ranged from $\sim 4.6$ to $26 \,\mu \rm W\, mol^{-1}.$ }  \gr{For less abundant elements where only weaker limits were required, 
we adopted smaller sample masses; here, the uncertainties in the molar heating rates  
ranged up to $870 \,\mu \rm W\, mol^{-1}.$} 
 
\section{Discussion}
 
As discussed by NFM18, the molar heating rate for a sample immersed in the dewar is
$$H_{\rm A} = 2\,R(T_{\rm DM}-T_{\rm LN2}) \,n_{\rm DM} {\bar v}\,\sigma^{\rm A}_{\rm 300\,K} {\bar f}_{\rm KE} $$
$$= 91 n_{14} (m_{\rm DM}/m_{\rm p})^{-1/2} (\sigma^{\rm A}_{\rm 300\,K}/{\rm mb}){\bar f}_{\rm KE} \,\rm \mu W\, mol^{-1}, \eqno(1)$$  
where $T_{\rm DM} \sim 300 \,\rm K$ is the temperature of the dark matter (assumed to be in equilbrium with the laboratory),
$T_{\rm LN2} = 77\,\rm K$ is the temperature of the dewar contents, $n_{\rm DM}(R_\earth) = 10^{14}\,n_{\rm 14}\,\rm cm^{-3}$ 
is the number density of HIDM at the
Earth's surface, ${\bar v} = (8 kT_{\rm DM}/(\pi m_{\rm DM}))^{1/2}$ is the mean speed of the HIDM, $m_{\rm DM}$ 
is the HIDM particle mass,
$\sigma^{\rm A}_{\rm 300\,K}$ is the average 
momentum transfer cross-section for collisions between 300~K HIDM and nucleus A, 
mb denotes the millibarn 
($1\,\rm mb = 10^{-27}\,\rm cm^2$) and
${\bar f}_{\rm KE}$ is the mean fraction of the kinetic energy transferred per collision.
For collisions that are isotropic in the center-of-mass frame, the average
fractional kinetic energy transfer is 
$$f_{\rm KE} = {2\,m_{\rm DM} m_{\rm A} \over (m_{\rm DM} + m_{\rm A})^2}, \eqno(2)$$
where $m_{\rm A}$ is the mass of the nucleus.
For the case where the interaction cross-section, $\sigma_{\rm v}^{\rm A}$, depends on velocity,
the appropriate average for use in equation (1) is 
$$\sigma_{\rm T}^{\rm A} = {\int f_{\rm MB}(v) v^3 \sigma_{\rm v}^{\rm A} dv \over f_{\rm MB}(v) v^3 dv},\eqno(3)$$
(NFM18), where $f_{\rm MB}(v)$ is the Maxwell-Boltzmann distribution of particle speeds at temperature $T$.

In Table 2 (column 4), we list the values of $\sigma^{\rm A}_{\rm 300\,K}\,n_{14}$ 
implied by the molar heating rates that were measured in our experiment.  The values given here were computed 
for an assumed 
particle mass of $2\,m_{\rm p}$.  Results for other values of $m_{\rm DM}$ may be obtained by 
multiplying the cross-sections by  
$m_{\rm DM}^{-1/2}(m_{\rm A}+m_{\rm DM})^2/(m_{\rm A}+2\,m_{\rm p})^2$.
If $m_{\rm DM} \ll m_{\rm A}$ and $m_{\rm A} \gg m_{\rm p}$, this factor is well approximated by $m_{\rm DM}^{-1/2}.$ 
\gr{These results are presented in graphical form in Figure 3, where upper limits (99$\%$ confidence) are plotted as 
a function of nucleon number for each element tested.  Limits for the most abundant isotope of each element are 
shown in red, with results for other stable isotopes in blue.  Here, we have used the natural isotopic abundances of 
Rosman \& Taylor (1998).}

\begin{table}
\scriptsize
\caption{\ma{Interaction cross-sections$^a$}}
\vskip 0.1 true in
\begin{tabular}{lccrrc}
\hline

Nucleus	& Fraction by 	& Molar fraction 	& $\sigma^{A}_{300\,K} n_{14}\phantom{00000}$ & $f_{\rm A}\sigma^{A}_{300\,K} n_{14}\phantom{00000}$  & Cum. sum\\
		& wt.\ in crust	& in crust, $f_{\rm A}$	& & & $\sum f_{\rm A}\sigma^{A}_{300\,K} n_{14}$ \\
 				&			&				 & 	$(\rm cm^2) \phantom{00000000}$	& $(\rm cm^2) \phantom{00000000}$ & (mb) \\
\hline
O         &  $4.66 \times 10^{-1}$	& $6.01 \times 10^{-1}$ 		& $(-5.38 \pm 7.99)\times 10^{-28}$ & $(-3.23 \pm 4.80)\times 10^{-28}$ & $  -0.323 \pm   0.480 $ \\
Si        &  $3.03 \times 10^{-1}$	& $2.23 \times 10^{-1}$ 		& $(+5.68 \pm 7.07)\times 10^{-28}$ & $(+1.27 \pm 1.58)\times 10^{-28}$ & $  -0.197 \pm   0.506 $ \\
Al        &  $7.74 \times 10^{-2}$	& $5.92 \times 10^{-2}$ 		& $(-5.93 \pm 6.96)\times 10^{-28}$ & $(-3.51 \pm 4.12)\times 10^{-29}$ & $  -0.232 \pm   0.507 $ \\
H         &  $1.50 \times 10^{-3}$	& $3.07 \times 10^{-2}$ 		& $(-1.19 \pm 1.77)\times 10^{-28}$ & $(-3.66 \pm 5.44)\times 10^{-30}$ & $  -0.235 \pm   0.507 $ \\
Na        &  $2.57 \times 10^{-2}$	& $2.30 \times 10^{-2}$ 		& $(+0.25 \pm 1.66)\times 10^{-27}$ & $(+0.58 \pm 3.82)\times 10^{-29}$ & $  -0.229 \pm   0.509 $ \\
Ca        &  $2.94 \times 10^{-2}$	& $1.52 \times 10^{-2}$ 		& $(+2.49 \pm 3.20)\times 10^{-27}$ & $(+3.77 \pm 4.85)\times 10^{-29}$ & $  -0.192 \pm   0.511 $ \\
K         &  $2.87 \times 10^{-2}$	& $1.51 \times 10^{-2}$ 		& $(-3.24 \pm 4.20)\times 10^{-27}$ & $(-4.90 \pm 6.36)\times 10^{-29}$ & $  -0.241 \pm   0.515 $ \\
Mg        &  $1.35 \times 10^{-2}$	& $1.15 \times 10^{-2}$ 		& $(-15.96 \pm 4.87)\times 10^{-28}$ & $(-18.31 \pm 5.58)\times 10^{-30}$ & $  -0.259 \pm   0.515 $ \\
Fe        &  $3.09 \times 10^{-2}$	& $1.14 \times 10^{-2}$ 		& $(+2.41 \pm 2.83)\times 10^{-27}$ & $(+2.75 \pm 3.23)\times 10^{-29}$ & $  -0.232 \pm   0.516 $ \\
C         &  $3.24 \times 10^{-3}$	& $5.57 \times 10^{-3}$ 		& $(-2.01 \pm 3.42)\times 10^{-28}$ & $(-1.12 \pm 1.90)\times 10^{-30}$ & $  -0.233 \pm   0.516 $ \\
Ti        &  $3.12 \times 10^{-3}$	& $1.34 \times 10^{-3}$ 		& $(-2.04 \pm 5.75)\times 10^{-27}$ & $(-2.74 \pm 7.73)\times 10^{-30}$ & $  -0.235 \pm   0.516 $ \\
F         &  $6.11 \times 10^{-4}$	& $6.64 \times 10^{-4}$ 		& $(+6.54 \pm 8.40)\times 10^{-28}$ & $(+4.34 \pm 5.57)\times 10^{-31}$ & $  -0.235 \pm   0.516 $ \\
S         &  $9.53 \times 10^{-4}$	& $6.13 \times 10^{-4}$ 		& $(-1.38 \pm 3.30)\times 10^{-27}$ & $(-0.85 \pm 2.03)\times 10^{-30}$ & $  -0.236 \pm   0.516 $ \\
P         &  $6.65 \times 10^{-4}$	& $4.43 \times 10^{-4}$ 		& $(-0.44 \pm 1.05)\times 10^{-26}$ & $(-1.95 \pm 4.67)\times 10^{-30}$ & $  -0.238 \pm   0.516 $ \\
Cl        &  $6.40 \times 10^{-4}$	& $3.73 \times 10^{-4}$ 		& $(+0.37 \pm 2.41)\times 10^{-27}$ & $(+1.38 \pm 8.98)\times 10^{-31}$ & $  -0.238 \pm   0.516 $ \\
Mn        &  $5.27 \times 10^{-4}$	& $1.98 \times 10^{-4}$ 		& $(+10.01 \pm 6.26)\times 10^{-27}$ & $(+1.98 \pm 1.24)\times 10^{-30}$ & $  -0.236 \pm   0.516 $ \\
N         &  $8.30 \times 10^{-5}$	& $1.22 \times 10^{-4}$ 		& $(-4.66 \pm 2.17)\times 10^{-28}$ & $(-5.70 \pm 2.66)\times 10^{-32}$ & $  -0.236 \pm   0.516 $ \\
Ba        &  $6.68 \times 10^{-4}$	& $1.00 \times 10^{-4}$ 		& $(+0.31 \pm 1.10)\times 10^{-25}$ & $(+0.31 \pm 1.10)\times 10^{-29}$ & $  -0.233 \pm   0.516 $ \\
Sr        &  $3.16 \times 10^{-4}$	& $7.44 \times 10^{-5}$ 		& $(+1.59 \pm 5.71)\times 10^{-26}$ & $(+1.18 \pm 4.25)\times 10^{-30}$ & $  -0.231 \pm   0.516 $ \\
Li        &  $2.20 \times 10^{-5}$	& $6.54 \times 10^{-5}$ 		& $(+0.21 \pm 3.65)\times 10^{-27}$ & $(+0.13 \pm 2.38)\times 10^{-31}$ & $  -0.231 \pm   0.516 $ \\
Zr        &  $2.37 \times 10^{-4}$	& $5.36 \times 10^{-5}$ 		& $(+0.28 \pm 1.02)\times 10^{-25}$ & $(+1.52 \pm 5.45)\times 10^{-30}$ & $  -0.230 \pm   0.516 $ \\
B         &  $1.70 \times 10^{-5}$	& $3.25 \times 10^{-5}$ 		& $(+0.33 \pm 5.81)\times 10^{-27}$ & $(+0.11 \pm 1.89)\times 10^{-31}$ & $  -0.230 \pm   0.516 $ \\
Rb        &  $1.10 \times 10^{-4}$	& $2.66 \times 10^{-5}$ 		& $(+0.16 \pm 2.87)\times 10^{-25}$ & $(+0.43 \pm 7.62)\times 10^{-30}$ & $  -0.229 \pm   0.516 $ \\
V         &  $5.30 \times 10^{-5}$	& $2.15 \times 10^{-5}$ 		& $(-1.40 \pm 3.34)\times 10^{-26}$ & $(-3.00 \pm 7.18)\times 10^{-31}$ & $  -0.230 \pm   0.516 $ \\
Zn        &  $5.20 \times 10^{-5}$	& $1.64 \times 10^{-5}$ 		& $(+0.51 \pm 9.02)\times 10^{-27}$ & $(+0.08 \pm 1.48)\times 10^{-31}$ & $  -0.230 \pm   0.516 $ \\
Cr        &  $3.50 \times 10^{-5}$	& $1.39 \times 10^{-5}$ 		& $(+0.70 \pm 2.53)\times 10^{-26}$ & $(+0.98 \pm 3.51)\times 10^{-31}$ & $  -0.230 \pm   0.516 $ \\
Ce        &  $6.57 \times 10^{-5}$	& $9.68 \times 10^{-6}$ 		& $(+0.40 \pm 7.12)\times 10^{-26}$ & $(+0.39 \pm 6.89)\times 10^{-31}$ & $  -0.230 \pm   0.516 $ \\
\hline 
\end{tabular}
\tablenotetext{a}{In cases where multiple nuclei were present for a given run (either because the 
sample material was a compound rather than a pure element or because multiple materials were tested at the same time), the 
values of $\sigma^{\rm A}_{\rm 300\,K}\,n_{14}$ were computed under the assumption that there 
is no heating associated with the other elements that were present.  This assumption, of course, makes the implied
upper limits conservative.}

\end{table}
\begin{figure}
\includegraphics[width=16 cm]{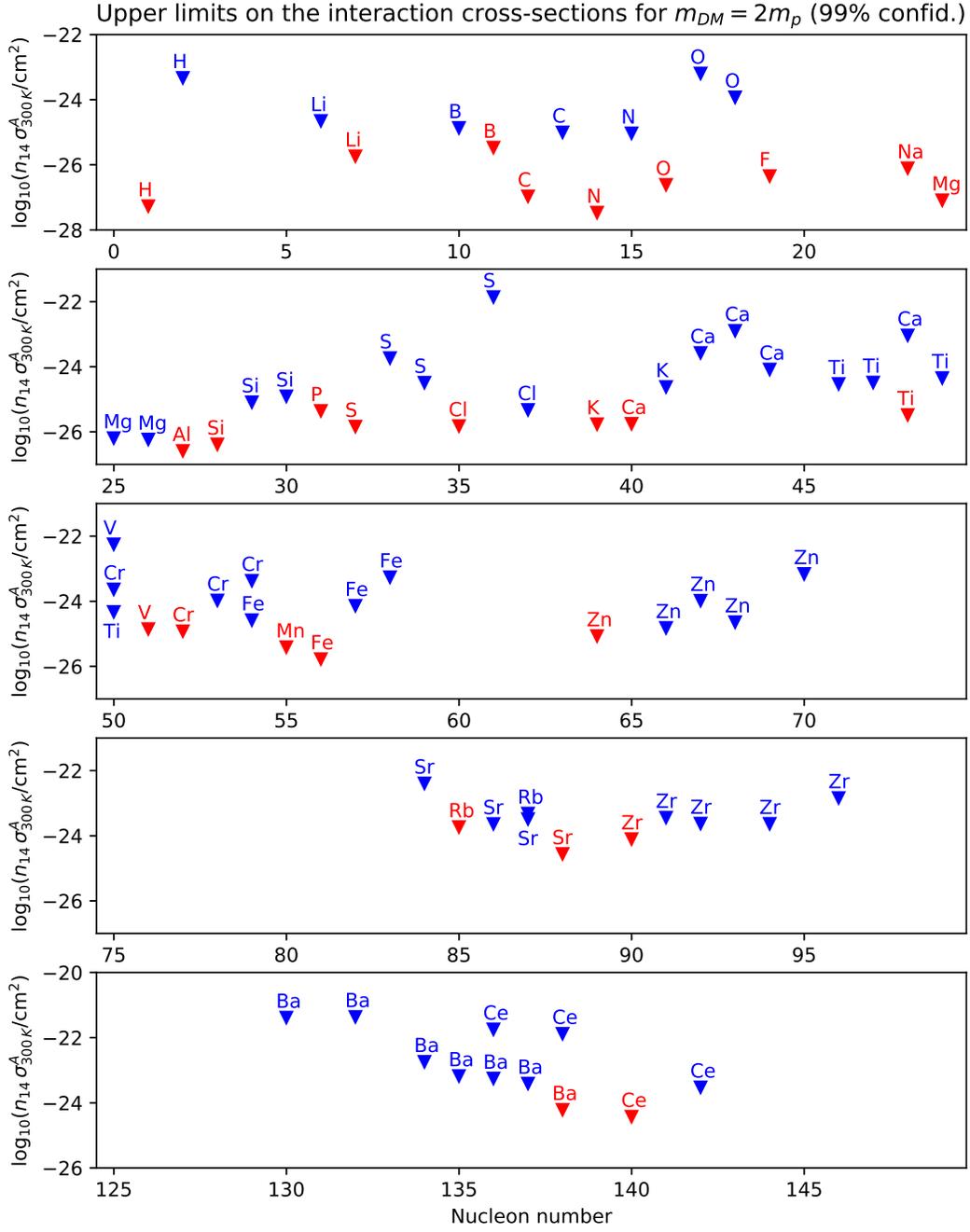}
\caption{Upper limits (99$\%$ confidence) on the interaction cross-sections for $m_{\rm DM} = 2\,m_{\rm p}$.}
\end{figure}

Based upon these results for $\sigma^{\rm A}_{\rm 300\,K}\,n_{14}$, we may compute the average value for material in the
crust:
$$\sigma^{\rm cr}_{\rm 300\,K} = \sum_A f_{\rm A} \sigma^{\rm A}_{\rm 300\,K}, \eqno(4)$$
where $f_{\rm A}$ is the molar fraction of atom A within Earth's crust.  In Table 2, we also list the elemental fractions
by weight in the crust (Wedepohl 1995), the corresponding molar fractions $f_{\rm A}$, the individual terms
contributing to equation (4), and the cumulative partial sum of those terms.  
The elements in Table 2 are listed in declining order of $f_{\rm A}$. \gr{As indicated by the cumulative partial sums, 
our measurements for the most abundant elements (particularly O and Si) dominate the uncertainties in
$\sigma^{\rm cr}_{\rm 300\,K}$; this justifies our use of smaller sample masses for the less abundant elements.}
Including all 27 elements with $f_{\rm A} \simgt 10^{-5}$,
we obtain 
$$\sigma^{\rm cr}_{\rm 300\,K} \le \gr{1.32} \, n_{\rm 14}^{-1} 
(m_{\rm DM}/2\,m_{\rm p})^{-1/2}\,\rm mb\,\,(3 \sigma). \eqno(5)$$
Here, we have assumed that $m_{\rm DM}$ is much smaller than $m_{\rm A}$ for typical crustal elements.

This result may be compared with a {\it lower} limit on the cross-section obtained by NFM18 from their consideration 
of heat transport through Earth's crust: 
$$\sigma^{\prime {\rm cr}}_{\rm 300\,K} \ge 51\,(m_{\rm DM}/2\,m_{\rm p})^{-1/2}
n_{\rm 14}\,\rm mb \,(3 \sigma). \eqno(6) $$ 
This constraint arises because HIDM particles contribute an additional thermal conductivity that is 
proportional to their mean-free path and therefore inversely proportional to the interaction cross-section.
Because observations of temperature gradients in boreholes place upper limits on any 
additional conductivity associated with HIDM (NFM18), equation (6) provides a lower limit on the cross-section.
The relevant cross-section in equation (6), which NFM18 denote 
$\sigma_{\rm 300 \, \rm K}^{\prime {\rm cr}}$ (with the use of a prime to distinguish it), 
has a different velocity-weighting than that appearing in equation (5), 
denoted $\sigma_{\rm 300 \, \rm K}^{\rm cr}$ without a prime.  

If the interaction cross-section is independent of 
velocity, the two cross-sections are identical: $\sigma_{\rm 300 \, rm K}^{\prime {\rm cr}} = 
\sigma_{\rm 300 \, \rm K}^{\rm cr}.$   In this case, the upper limit on $\sigma_{\rm 300 \, \rm K}^{\rm cr}$ (equation 5)
can be combined with the lower limit on $\sigma_{\rm 300 \, \rm K}^{\prime {\rm cr}}$ (equation 6) to obtain a conservative
upper limit on the particle density 
at the Earth's surface\footnote{If the interaction cross-section has a velocity dependence of the form
$\sigma_{\rm v}^{\rm A} \propto v^{-j}$, NFM18 showed
that $\sigma_{\rm 300 \, \rm K}^{\prime {\rm cr}}/  \sigma_{\rm 300 \, \rm K}^{{\rm cr}} 
= 4 / [\Gamma(3+\onehalf j) \Gamma(3-\onehalf j)],$ where $\Gamma$ denotes the usual gamma function.  
In this general case, our limit on the particle density becomes
$$\rm n_{14} \le 0.36 / [\Gamma(3+\onehalf j) \Gamma(3-\onehalf j)]^{1/2}$$
The right hand side of equation (7) yields values of \gr{0.16, 0.15, 0.13, 0.10, 0.066} respectively for $j$ 
= 0, 1, 2, 3, and 4.}: 
$$n_{\rm DM}(R_\earth)  \le 1.6 \times 10^{13}\,\rm cm^{-3} \eqno(7)$$

\begin{figure}
\includegraphics[width=17 cm]{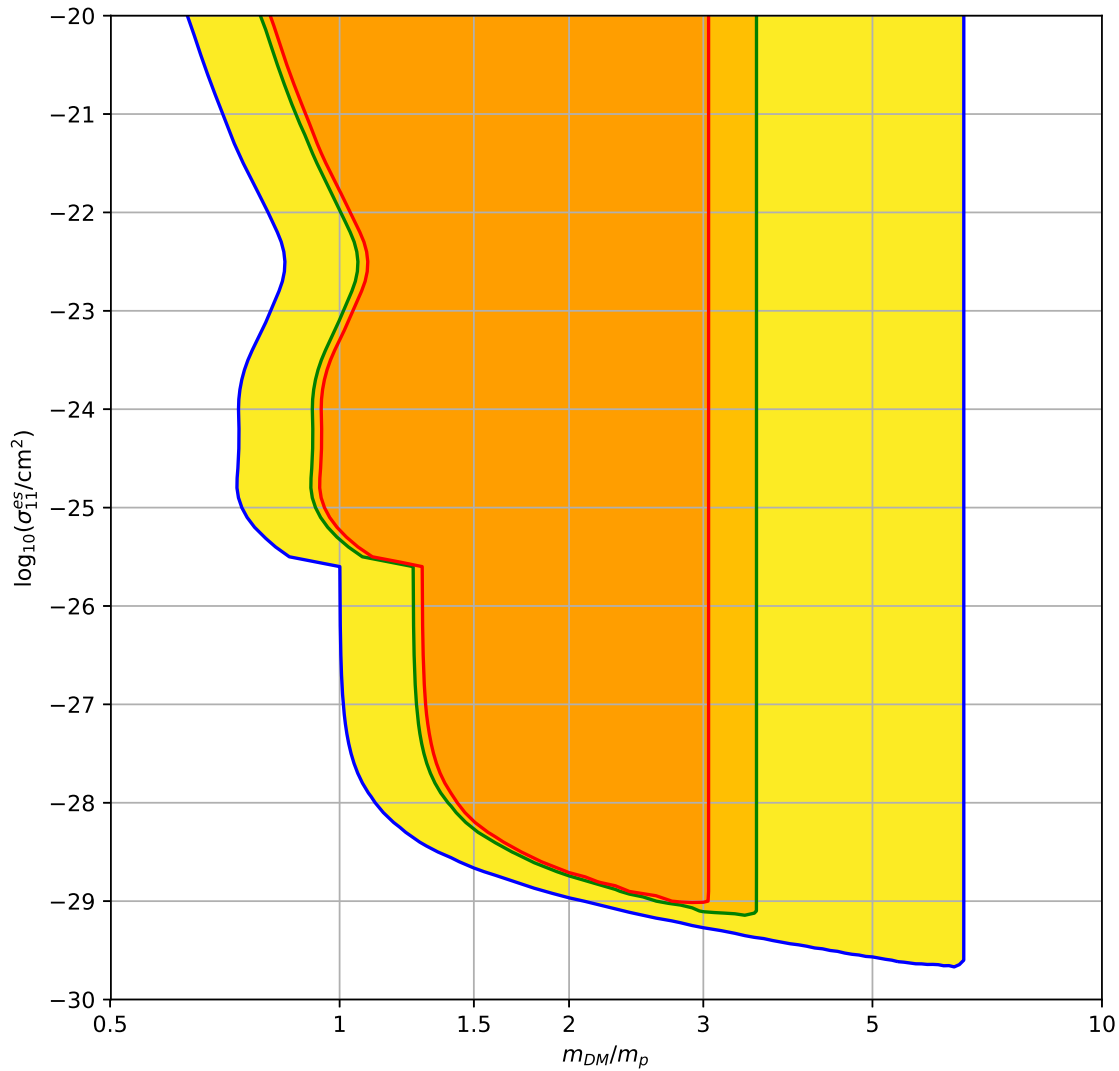}
\caption{Region of parameter space excluded by our experiment at the 3 $\sigma$ level).  Dark orange region: excluded if the
cross-section is independent of velocity.  Dark and light orange regions: excluded if the
cross-section is proportional to $v^{-4}$.  Yellow: region that could be explored with a factor 1000 improvement
in experimental sensitivity.}
\end{figure}

These upper limits on the HIDM particle density place strong constraints on the particle mass and cross-section.  
The HIDM densities, $n_{\rm DM}(R_\earth)$, predicted by NFM18 at the surface of the Earth depend upon two parameters: 
the particle mass, $m_{\rm DM}$, and the mean cross-section for escaping particles, $\sigma^{\rm es}_{11}$.  The latter 
parameter determines the location of the LSS, which in turn determines the temperature at the LSS
and thus the rate at which captured HIDM are lost by Jeans escape.  Here, the interaction cross-section 
applies to a velocity of $11.2\,\rm km\,s^{-1}$ \gr{(the escape velocity from the Earth), which we 
indicate notationally by the subscript 11.}  If the LSS is in the
atmosphere, then $\sigma^{\rm es}_{11}$ is the appropriate average for a medium of atmospheric 
composition, $\sigma^{\rm atm}_{11}$. 
This condition applies if $\sigma^{\rm atm}_{11} \ge 25~$mb; otherwise, 
the LSS lies \gr{beneath the Earth's surface.  If the LSS lies in the crust, 
$\sigma^{\rm es}_{11}$ has the value $\sigma^{\rm cr}_{11}$.}  
In Figure 4, which is
based on NFM18 Figure 6, we show the region of parameter space that is now excluded by our experiment at the 3$\,\sigma$ level.  
For a velocity-independent cross-section, the dark orange region bounded by the red coutour is excluded, while
for cross-section proportional to $v^{-4}$, the slightly larger region bounded by the green contour is ruled out.
A particle mass of $2\,m_{\rm p}$ is excluded unless $\sigma^{\rm es}_{11} \le 2 \times 10^{-29}\,\rm cm^{2}$; 
\gr{and for $\sigma^{\rm es}_{11} \ge 10^{-28}\,\rm cm^{2}$, particle masses between 1.4 and $3\,m_{\rm p}$ are 
ruled out.}
The blue contour shows how the region explored would expand if the sensitivity of the experiment could be improved by
a factor 1000 (see section 5 below).

The dark orange region in Figure 4 is now excluded for almost any plausible variation of the interaction cross-section from  
one nucleus to another.  The one caveat is that we assume here that none of the remaining nuclei that have not been tested 
in our experiment has an extremely large cross-section for scattering with HIDM.   Because we have tested all elements 
with a molar fraction $\simgt 10^{-5}$ in the crust, any single element among those that have not been tested
would have to exhibit a cross-section greater than $5.1 \times 10^6$~mb ($5.1 \times 10^{-21} \, \rm cm^2$) to permit 
$n_{14} \ge 1$.  Collectively, the untested nuclei account for a combined molar fraction of \re{$5.9 \times 10^{-5}$}.  
Thus, if multiple untested nuclei had enhanced cross-sections for interaction with HIDM, cross-sections of at 
least \re{$8.6 \times 10^5$~mb ($8.6 \times 10^{-22} \, \rm cm^2$)} would be needed to permit $n_{\rm 14} \ge 1.$

\section{\gr{Summary and future prospects}} 
 
\noindent 1) To probe dark matter that interacts with baryons,
we have performed a simple and inexpensive tabletop experiment to search for anomalous heating in 24 materials
cooled to 77~K in a LN2 dewar.   The experiment involved a differential measurement, 
in which daily weight measurements -- with a fractional precision $\sim 10^{-5}$ -- were used to compare the
LN2 evaporation rates between alternating runs of duration $\sim 6$ days conducted with and without each sample material. 

\noindent 2) Our experiment placed upper limits in the range 
0.40 -- 0.57 mW on any heating rate associated with the interaction of DM with the tested samples, which ranged in
mass from 59 to 2192~g.  Upper limits on the molar heating rates, which ranged from $4.8\,\mu$W~mol$^{-1}$ to 
$3.9\,$mW~mol$^{-1}$, were thereby obtained for the 27 elements with molar
fractions greater than $\sim 10^{-5}$ in Earth's crust:
H, Li, B, C, N, O, F, Na, Mg, Al, Si, P, S, Cl, K, Ca, Ti, V, Cr, Mn, Fe, Zn, Rb, Sr, Zr, Ba, and Ce.  These constraints
place upper limits on the cross-sections for the interaction of DM with those elements, which are plotted in Figure 3  
for each stable isotope.
 
\noindent 3) For material of the typical elemental composition in the crust, our measurements imply a $3 \sigma$ 
upper  limit of $1.32 \, n_{\rm 14}^{-1} (m_{\rm DM}/2\,m_{\rm p})^{-1/2}\,\rm mb$ on the mean 
cross-section for scattering with thermal HIDM at 300~K.  In combination with a lower limit on the cross-section
that was obtained (NFM18) by considering the measured thermal conductivity of the crust, this upper limit 
implies that the density of HIDM at the Earth's surface must be less than $1.6 \times 10^{13}\, \rm cm^{-3}$. Figure 4
shows the region of parameter space that is excluded by this limit on the HIDM density.

While the current experimental design has proven adequate in providing 
strong constraints on the particle properties, significant improvements could potentially be achieved.  The use of 
liquid helium (LHe) in place of LN2 would provide enhanced sensitivity to anomalous heating,
because LHe possesses a specific latent heat of
vaporization that is a factor of 20 smaller than that of LN2.  The use of larger dewars, along
with larger sample sizes and longer run durations, could provide more sensitive determinations of the 
boil-off rate if systematic changes in the dewar heat leak could be eliminated with the use of 
more robust dewars.  Such enhancements might plausibly improve the sensitivity of the 
experiment by three orders of magnitude, allowing the yellow region in Figure 4 to be probed.

\begin{acknowledgements}

We thank C.\ F.\ McKee for helpful comments on the manuscript.

\end{acknowledgements}

\end{document}